# FastSIR Algorithm: A Fast Algorithm for simulation of epidemic spread in large networks by using SIR compartment model

Nino Antulov-Fantulin[a,1,*], Alen Lančić[b], Hrvoje Štefančić[c], Mile Šikić[d,e,2,**]

[a]*Division of Electronics, Laboratory for Information Systems,*
*Rudjer Bošković Institute, Zagreb, Croatia*
[b]*Faculty of Science, Department of Mathematics, University of Zagreb, Zagreb, Croatia*
[c]*Theoretical Physics Division, Rudjer Bošković Institute, Zagreb, Croatia*
[d]*Faculty of Electrical Engineering and Computing, Department of Electronic Systems and Information Processing,*
*University of Zagreb, Croatia*
[e]*Bioinformatics Institute, A*STAR, Singapore, Republic of Singapore*

**Abstract**

The epidemic spreading on arbitrary complex networks is studied in SIR (Susceptible Infected Recovered) compartment model. We propose our implementation of a Naive SIR algorithm for epidemic simulation spreading on networks that uses data structures efficiently to reduce running time. The Naive SIR algorithm models full epidemic dynamics and can be easily upgraded to parallel version. We also propose novel algorithm for epidemic simulation spreading on networks called the FastSIR algorithm that has better average case running time than the Naive SIR algorithm. The FastSIR algorithm uses novel approach to reduce average case running time by constant factor by using probability distributions of the number of infected nodes. Moreover, the FastSIR algorithm does not follow epidemic dynamics in time, but still captures all infection transfers. Furthermore, we also propose an efficient recursive method for calculating probability distributions of the number of infected nodes. Average case running time of both algorithms has also been derived and experimental analysis was made on five different empirical complex networks.

## 1. Introduction

Complex networks represent structure of communication networks [1] [2] or social contact interactions [3] [4] very well. Therefore, it is reasonable to study computer virus propagation or epidemic spreading on complex networks [5] [6] [7]. Modeling the spread of an epidemic in a population is usually done by dividing individuals of the population into subdivision with some common characteristic features called compartments. The SIR model is a good model for many infectious diseases where each individual in a population can be in one of three different compartments. Those who are susceptible to the disease are in the Susceptible compartment, those who are infected and can transmit the disease to others are in the Infected compartment and those who have recovered and are immune and those who are removed from population are in the Recovered compartment. Some infectious diseases are described with models that have different number of compartments like SIS model (Susceptible Infected Susceptible) where individuals can not have long lasting immunity and therefore Recovered compartment does not exist.

[*]Corresponding author.
[**]Corresponding author.
  *Email addresses:* `nino.antulov@irb.hr` (Nino Antulov-Fantulin), `alen@student.math.hr` (Alen Lančić), `shrvoje@thphys.irb.hr` (Hrvoje Štefančić), `mile.sikic@fer.hr` (Mile Šikić)
[1]P.O.B. 180, HR-10002, Zagreb
[2]30 Biopolis Street, 07-01 Matrix, Singapore 138671

Different mathematical models have been used to study epidemic spreading. Under the assumption of homogeneous mixing among individuals inside different compartments (Kermack-McKendrick model) differential equations can be applied to understand epidemic dynamics [12]. Contact network epidemiology applies bond percolation on random graphs (not on arbitrary structure) to model epidemic spreading on heterogeneous population (epidemic dynamics is neglected) [13] [14] [6]. Small world network property [8] and scale-free network property [9] [10] have great impact on epidemic spreading outcome. Scale-free complex networks exhibit no epidemic threshold below which the infection cannot produce epidemic outbreak (endemic state) in SIS model [11].

Realistic epidemic simulations (EpiFast [16], EpiSims [17] and EpiSindemics [18]) have become very important application of high-performance computing in epidemic predictions. These are just a few examples of parallel algorithms that can be used in public health studies. Some studies [19] [20] used contact network models between urban cities (cities are connected through airline transportation network) and homogeneous mixing model inside urban cities and examined influence of interventions (antiviral drugs and containments) to worldwide spread of pandemic.

Recently, the phase diagrams of epidemic spreading with the SIR model on complex networks were introduced as a useful tool for epidemic spreading analysis [21]. To predict expected value of the number of infected nodes in epidemic spreading on arbitrary network, we should repeat simulations till standard deviation of the number of the infected nodes sufficiently reduces (in unimodal part of the epidemic phase diagram) or converges to nonzero value (in bimodal part of the epidemic phase diagram). Due to the high number of iterations needed to predict expected epidemic outcome, average case running time of the epidemic simulation algorithm should be as low as possible. In this paper we describe our implementation of a Naive SIR algorithm to simulate epidemic spreading (SIR model) on arbitrary network structure with low average case running time. Our implementation of the Naive SIR algorithm uses data structures efficiently to reduce running time. This algorithm models full dynamics of epidemic spreading and can be upgraded easily to parallel version. Main contribution of this paper is a novel algorithm called the FastSIR algorithm which uses probability distributions of the number of infected nodes to speed up recovery of infected nodes to one discrete step during simulation of epidemic spreading to reduce running time. Moreover, the FastSIR algorithm does not follow epidemic dynamics in time, but still captures all infection transfers.

In section 2 we formally define an epidemic simulation problem and other concepts used in this paper. Section 3 describes our implementation of Naive SIR algorithm along with the running time and the space complexity analysis. In section 4 we describe our novel FastSIR algorithm along with the running time and the space complexity analysis. Correctness of the FastSIR algorithm was proven with four steps of equality. We also described how to efficiently implement probability distributions of the number of infected nodes by a recursive method. In section 5 we described results of the performance profiling and analysis of our algorithms on five empirical complex networks. In sections 6 and 7 we described possible applications of our algorithms along with discussion of results and conclusion.

## 2. Epidemic simulation problem

We define contact-network as undirected and non-weighted graph $G(N, L)$ ($N$-set of nodes, $L$-set of links). Link $(u, v)$ exists only if two nodes $u$ and $v$ are in contact during epidemic time. We also assume that the contact-network during epidemic process is static one, but the algorithms and other results can also be applied even for dynamic networks. To simulate epidemic propagation through contact-network, we use standard stochastic SIR model. In this model each node at some time can be in one of the following states: susceptible (S), infected (I) and recovered (R). A discrete time model is used. Time needed for the epidemic to stop spreading defines one epidemic simulation. At the beginning of each epidemic simulation all nodes from graph $G$ are in susceptible state except arbitrary set of nodes, which are initially infected, with some epidemic parameters $p$ and $q$. These initial conditions we denote with letter $\lambda$. Epidemic parameter $p$ is a probability that an infected node $u$ infects adjacent susceptible node $v$ in one discrete time step. Epidemic parameter $q$ is a probability that an infected node recovers in one discrete time step. At the end of an epidemic simulation all nodes can be in one of two following states: susceptible or recovered. Epidemic simulation on $(G, \lambda)$ is a random instance of epidemic stochastic process $ESP(G, \lambda)$. Epidemic simulations are mutually independent. Let $X$ be a random variable that measures a number of infected nodes of epidemic stochastic process $ESP(G, \lambda)$. Given a $ESP(G, \lambda)$, the epidemic simulation problem is to compute first $n$ moments of random variable $X$. We will also often use the notation $\mathbb{P}(X_n = k)$, which denotes the probability that the infected node infects $k$ neighbours out of total $n$ susceptible neighbours in the limit of the time. This probability has the following analytical form [21] in the SIR model:

$$\mathbb{P}(X_n = k) = q \binom{n}{k} \sum_{l=0}^{k} \binom{k}{l} (-1)^l \frac{(1-p)^{n-k+l}}{1 - (1-q)(1-p)^{n-k+l}}.$$

## 3. The Naive SIR algorithm

In standard algorithm for SIR model, an infected node tries to infect its neighbors sequentially. For each neighboring node a pseudo random number between 0 and 1 is calculated. If the number is smaller or equal to $p$ value, the neighboring node is infected. At the end we check if the node recovers according to a new pseudo random number and $q$ parameter. In this paper we call this algorithm the Naive SIR algorithm.

In our implementation (see Algorithm 1) we use a queue for the set of infected nodes $I$ and an array structure $S$ for indication of susceptible nodes. If the array value of particular node is "1" that node is susceptible. Vice versa, the node is infected or recovered. The network was represented using an adjacency list.

*3.1. Time and space complexity analysis of the Naive SIR algorithm*

Here, we examine the average case running time and space complexity of the Naive SIR algorithm. For order of growth of average case running time algorithm analysis we use standard big-$O$ notation (asymptotic upper bound within a constant factor) [30].

**Algorithm 1** The Naive SIR algorithm
---
**Input:** $(G, \lambda)$ where $G$ is contact network and $\lambda$ represents the initial conditions. Initial conditions consist of $p$, $q$, $I$ a queue of initally infected nodes and $S(v)$ is an array indicator of susceptible nodes.
**Output:** array indicator of recovered nodes $R(v)$
**while** $I$ is not empty **do**
  **dequeue**(u, I)
  **for each** contact $v$ of node $u$ **do**
    **if** $S(v)$ is equal to 1 **then**
      let transmission of infection $u \Rightarrow v$ occur with probability $p$
      **if** $u \Rightarrow v$ does occur **then**
        update $S(v)$ and $R(v)$
        enqueue $(v, I)$
      **end if**
    **end if**
  **end for**
  update state of $u$ from infected to recovered with probability $q$
  **if** $u$ is not recovered **then**
    enqueue(u, I)
  **end if**
**end while**
output $R(v)$

---

The average case running time of the Naive SIR algorithm $\overline{T_c}(\mathbb{E}[X], \overline{k}, q)$ is equal to:

$$\overline{T_c}(\mathbb{E}[X], \overline{k}, q) = O\left(\frac{\mathbb{E}[X]\overline{k}}{q}\right) \tag{1}$$

where $\mathbb{E}[X]$ denotes total expected number of infected nodes and $\overline{k}$ denotes average degree.

To explain this statement, let us start with the case of one infected node with $k$ neighbors. In one cycle it tries to transmit infection to each of its neighbors. The run-time calculation cost of that is proportional to $k$. At the end of each cycle a random number is compared with $q$. If the number is greater than $q$ the node is moved to set of recovered nodes. Total running time cost $T_c^i$ for some infected node $v_i$ is sum of costs over all time steps where node $v_i$ was infected. Hence, it can been seen that number of cycles the node is in infected state is a sample from geometric distribution with expectation $1/q$. Because of that, total average running time for one infected node is $\overline{T_c^1} = O(k/q)$. Let $\mathbb{E}[X]$ be the expected number of infected nodes in the network. Because main while loop of the Naive SIR algorithm executes sequentially total average case running time $\overline{T_c}$ is sum of $\overline{T_c^i}$ for all infected nodes $v_i$. The sum $\overline{T_c} = \overline{T_c^1} + \overline{T_c^2} + ... + \overline{T_c^n}$ has $\mathbb{E}[X]$ terms. Therefore average case running time $\overline{T_c}$ is $O(\mathbb{E}[X]\overline{k}\frac{1}{q})$.

For a network with cycles, it is difficult to analytically calculate the expected number of infected nodes, but we can calculate it for a regular m-arry tree. To that end, we will use $X_n$, random variable of a number of directly infected susceptible nodes by the infected node of degree $n$ [21]. It can be easily verified that $\mathbb{E}[X_n] = n\mathbb{E}[X_1] = n\mathbb{P}(X_1 = 1)$.

**Proposition 3.1.** *The average case running time of the Naive SIR algorithm $\overline{T_c}(\mathbb{E}[T_n], \overline{k})$ for a*

*m-arry tree of depth n is equal to:*

$$\overline{T_c}(\mathbb{E}[T_n], \overline{k}) = O(\mathbb{E}[T_n]\overline{k}),$$

where $T_n$ is a random variable that measures time needed for epidemic to stop spreading in regular m-arry tree of depth n and $\overline{k}$ denotes average degree.

In particular the expectation of $T_n$ satisfies the relation:

$$\mathbb{E}[T_n] \leqslant \frac{1}{q} \frac{[m\mathbb{P}(X_1 = 1)]^n - 1}{m\mathbb{P}(X_1 = 1) - 1}$$

where the expression $\frac{[m\mathbb{P}(X_1=1)]^n-1}{m\mathbb{P}(X_1=1)-1} = \mathbb{E}[X]$ is the expected total number of infected nodes [21].

PROOF. The upper bound on the expected value of a random variable $T_n$ is calculated by applying law of total expectation with partition $\{X_m = i : 0 \leqslant i \leqslant m\}$; in case $X_m = 0$ follows $\mathbb{E}[T_n|X_m = 0] = \mathbb{E}[T_0] = \frac{1}{q}$ and otherwise $\mathbb{E}[T_n|X_m = k] \leqslant \mathbb{E}[T_0] + \mathbb{E}\left[\max_{1\leqslant j\leqslant k} T_{n-1}^{(j)}\right]$ - the expected value of time needed for epidemic to stop spreading at root level plus the expected value of maximum of times needed for epidemic to stop spreading in each of $k$ subtrees with depth $n-1$, where $k = 1, 2, ..., m$

$$\mathbb{E}[T_n] = \sum_{i=0}^{m} \mathbb{E}[T_n|X_m = i]\mathbb{P}(X_m = i) \leqslant \mathbb{E}[T_0]\mathbb{P}(X_m = 0) +$$

$$+ \sum_{i=1}^{m} \left(\mathbb{E}[T_0] + \mathbb{E}\left[\max_{1\leqslant j\leqslant i} T_{n-1}^{(j)}\right]\right)\mathbb{P}(X_m = i) =$$

$$= \frac{1}{q} + \sum_{i=1}^{m} \mathbb{E}\left[\max_{1\leqslant j\leqslant i} T_{n-1}^{(j)}\right]\mathbb{P}(X_m = i) \leqslant \frac{1}{q} + \sum_{i=1}^{m} \mathbb{E}\left[\sum_{j=1}^{i} T_{n-1}^{(j)}\right]\mathbb{P}(X_m = i) =$$

$$= \frac{1}{q} + \mathbb{E}[T_{n-1}]\sum_{i=1}^{m} i \cdot \mathbb{P}(X_m = i) = \frac{1}{q} + \mathbb{E}[T_{n-1}]\mathbb{E}[X_m] =$$

$$= \frac{1}{q} + m\mathbb{E}[T_{n-1}]\mathbb{P}(X_1 = 1) \Rightarrow \mathbb{E}[T_n] \leqslant \frac{1}{q}\frac{[m\mathbb{P}(X_1 = 1)]^n - 1}{m\mathbb{P}(X_1 = 1) - 1}$$

The space complexity $S$ of the Naive SIR algorithm with respect to the number of links $L$ and the number of nodes $N$ is equal to:

$$S[L, N] \approx \underbrace{2L}_{G} + \underbrace{N}_{I} + \underbrace{N}_{S(v)} + \underbrace{N}_{R(v)} = 2L + 3N,$$

where the first term denotes space complexity of contact network $G$ (adjacency list), the second term denotes space complexity of a queue of infected nodes $I$, the third term denotes space complexity of an array indicator of susceptible nodes $S(v)$ and the last term denotes space complexity of an array indicator of recovered nodes $R(v)$. Note, that $S(v)$ and $R(v)$ can be implemented as a bitset structure to further reduce memory consumption.

In connected networks $L \geqslant N$ and then the space complexity $S$ of the Naive SIR algorithm is:

$$S[L, N] = O(L).$$

## 4. The FastSIR algorithm

The main goal of this article is to find a faster algorithm (see Algorithm 2) for determining the total number of infected nodes in epidemic, in which the stochastic simulation of epidemic spreading dynamics is not used explicitly. Looking at complexity of Algorithm 1, we can see that possible speed up of sequential version of the algorithm can be obtained only by reducing the $1/q$ part. Since we know how to calculate the probability distributions for the number of infected nodes [21], the idea is to choose that number from distribution. The probability that the infected node infects $k$ neighbours out of total $n$ susceptible neighbours in the limit of the time is:

$$\mathbb{P}(X_n = k) = q\binom{n}{k}\sum_{l=0}^{k}\binom{k}{l}(-1)^l \frac{(1-p)^{n-k+l}}{1-(1-q)(1-p)^{n-k+l}}. \qquad (2)$$

For the calculation of cumulative distribution $C_n(k) = \mathbb{P}(X_n \leq k)$, $p$, $q$ and $k$ should be known. These values can be calculated on the fly, but they can also be calculated in advance and saved on disk. In that case we do not need to repeat calculation for the same k values. Furthermore, since we use a few thousand simulations for each 3-tuple $p$, $q$, $k$, it is easy to see a benefit of the precalculated distributions. Distributions should be precalculated only once and can be used for several networks. The cost of calculation of a distributions for each $k$ up to some $k_{max}$ is proportional to $k_{max}^2$. However, the benefit of precalculation is evident in cases when it is necessary to run simulation using different starting parameters [28]. Furthermore, since contact social networks usually have $k_{max}$ up to tens of thousand, it is necessary to precalculate distribution once for all of them.

---
**Algorithm 2** The FastSIR algorithm
---
**Input:** $(G, \lambda, C)$ where $G$ is contact network and $\lambda$ represents the initial conditions, $C$ is cumulative distribution for $p$, $q$ and all $k$ values in the network. Initial conditions consist of $p$, $q$, $I$ a queue of initially infected nodes and $S(v)$ is an array indicator of susceptible nodes.
**Output:** array indicator of recovered nodes $R(v)$
**while** $I$ is not empty **do**
  **dequeue**$(u, I)$
  draw a pseudo random value $r$
  find from $C(k, p, q)$ a maximal value of $k_1$ such that $C(k_1, p, q) \leq r$, where $k_1$ is number of infected neighbors
  draw from $k$ neighbors $k_1$ nodes $w$
  **for each** $w$ **do**
    **if** $S(w)$ is equal to 1 **then**
      update $S(w)$ and $R(w)$
      enqueue $(w, I)$
    **end if**
  **end for**
**end while**
output $R(v)$
---

A distinction between simulations in the Naive SIR algorithm and the FastSIR algorithm is in the parameter that orders the execution of the simulation. For Naive SIR the simulation is ordered in (discrete) time: the simulation follows the dynamics of infection transfer as it unfolds in time. In the case of FastSIR, the parameter ordering the execution of the simulation is the parameter that we call the generation index. All infected nodes can be classified into generations according to number of infection transfers from the initially infected node. In particular, the initially infected node has a generation index 0, the nodes that it infects have the generation index 1 and so on. In FastSIR, the simulation starts from the initially infected node (generation 0) and using probability distributions for the number of infected nodes, nodes from generation 1 are determined and the node from generation 0 is recovered. In the n-th step of the simulation, starting from the nodes from generation $n-1$, the nodes from generation $n$ are determined using probability distributions for the number of infected nodes. Then the nodes from the generation $n-1$ are recovered and the simulation proceeds to the next step. Essentially, as a stochastic process, FastSIR captures all infection transfers happening in Naive SIR using different ordering (generation versus time).

*4.1. Correctness of the FastSIR algorithm*

To see correctness of the FastSIR algorithm (see Algorithm 2) we change Naive algorithm in a couple of steps that guarantee equality with respect to all infection transfers happening in Naive SIR process.

- First, all nodes infected directly by initially infected nodes can not recover nor infect their neighbors until the last of the initially infected nodes is recovered. Then, process is repeated so that all infected nodes in moment of recovery of the last initially infected node are defined as the initially infected nodes. It is clear that in this way the probability of infection of any neighbor of initially infected nodes directly by any of initially infected nodes remains unchanged. Since all probabilities of direct transfer of infection remain unchanged until the end of the algorithm, we conclude that this modification leaves a probability of infection of any node unchanged.

- The second step differs from the first step of modifying the Naive SIR algorithm in the way that all nodes except the initially infected node cannot recover nor infect their neighbors until the chosen initially infected node is recovered. Then we choose another node that plays the role of the initially infected node and repeat the process. Probabilities of direct infection of any of the neighbors of initially infected nodes are not changed because the probability of transmission of infection from each initially infected node to any of its neighboring nodes remained the same, and the order in which we have chosen initially infected nodes does not affect the probability of infection of susceptible neighbors of initially infected nodes.

- The third step is the reduction of all the steps of recovery of chosen initially infected node to a single step; if the initially infected node has $m$ susceptible neighbors, by using a distribution of a random variable of infection we can determine the realization of the number of infected nodes, and then realization of that number of infected nodes among the susceptible $m$. The probability of direct transmission of infection remains unchanged due to the construction of a random variable of infection.

- Fourth step uses the principle which we prove in the following proposition. It states that in the previous step, number $n$ of adjacent nodes can be taken instead of number $m$ of susceptible nodes, as long as only susceptible adjacent nodes are infected in the process.

**Proposition 4.1.** *Let node v have n neighbors of which $s_1, \ldots, s_m$ are susceptible and $i_{m+1}, \ldots, i_n$ cannot be infected, and let Y be a random variable of number of nodes infected directly by node v. Alternatively, let node v have the same n neighbours $s_1, \ldots, s_m, i_{m+1}, \ldots, i_n$ which are susceptible, and let Z be a random variable of number of nodes infected directly by node v among nodes $s_1, \ldots, s_m$. Then Y i Z are identically distributed random variables. Furthermore, in both instances node $s_i$ has the same probability of being directly infected by node v for all i, $1 \leqslant i \leqslant m$.*

PROOF. Let node $v$ have $m$ susceptible neighbours and degree $n$. Probability that $k$ out of $m$ susceptible neighbours end up being infected by node $v$ is obviously $\mathbb{P}(X_m = k)$. Probability that by infecting $n$ neighbours, out of which $m$ can be infected and $n - m$ can not, is obtained as follows:

Probability that $i$ predetermined nodes out of $n$ susceptible nodes become infected is $\mathbb{P}^*(X_n = i)$. We know that only $m$ out of $n$ nodes are actually susceptible, so we have to choose $i$ out of that $k$ nodes which are in the set of $m$ susceptible nodes, and remaining $i - k$ in the set of $n - m$ which can not be infected. That can be done in $\binom{n-m}{i-k}\binom{m}{k}$ different ways. Probability that in the end there are going to be $k$ infected nodes in the set of $m$ susceptible nodes is

$$\sum_{i=0}^{n} \binom{n-m}{i-k}\binom{m}{k} \mathbb{P}^*(X_n = i) = \sum_{i=k}^{n-m+k} \binom{n-m}{i-k}\binom{m}{k} \mathbb{P}^*(X_n = i) \tag{3}$$

The only thing left to do is compare these two expressions. We will use following relations:

$$\mathbb{P}(X_n = k) = q\binom{n}{k} \sum_{l=0}^{k} \binom{k}{l}(-1)^l \frac{(1-p)^{n-k+l}}{1-(1-q)(1-p)^{n-k+l}} \tag{4}$$

$$\mathbb{P}(X_n = k) = \binom{n}{k} \mathbb{P}^*(X_n = k) \tag{5}$$

$$\mathbb{P}(X_n = k) = q\binom{n}{k} \sum_{\mu=0}^{\infty} \left(1 - (1-p)^{1+\mu}\right)^k \left((1-p)^{1+\mu}\right)^{n-k} (1-q)^{\mu} \tag{6}$$

We have

$$\mathbb{P}^*(X_m = k) \stackrel{5,6}{=} q \sum_{\mu=0}^{\infty} (1 - (1-p)^{1+\mu})^k ((1-p)^{1+\mu})^{m-k}(1-q)^{\mu} =$$

$$= q \sum_{\mu=0}^{\infty} (1 - (1-p)^{1+\mu})^k ((1-p)^{1+\mu})^{m-k}(1-q)^{\mu} *$$

$$* \underbrace{\sum_{i=0}^{n-m} \binom{n-m}{i}(1 - (1-p)^{1+\mu})^i ((1-p)^{1+\mu})^{n-m-i}}_{=1} =$$

$$= q \sum_{\mu=0}^{\infty} \left(1 - (1-p)^{1+\mu}\right)^k \left((1-p)^{1+\mu}\right)^{m-k} (1-q)^{\mu} *$$

$$* \sum_{i=k}^{n-m+k} \binom{n-m}{i-k}\left(1-(1-p)^{1+\mu}\right)^{i-k}\left((1-p)^{1+\mu}\right)^{n-m-i+k} =$$

$$= q \sum_{\mu=0}^{\infty} (1-q)^{\mu} \sum_{i=k}^{n-m+k} \binom{n-m}{i-k}\left(1-(1-p)^{1+\mu}\right)^{i}\left((1-p)^{1+\mu}\right)^{n-i} =$$

$$= \sum_{i=k}^{n-m+k} \binom{n-m}{i-k} \underbrace{\sum_{\mu=0}^{\infty} q\left(1-(1-p)^{1+\mu}\right)^{i}\left((1-p)^{1+\mu}\right)^{n-i}(1-q)^{\mu}}_{=\mathbb{P}^*(X_n=i)} = \sum_{i=k}^{n-m+k} \binom{n-m}{i-k}\mathbb{P}^*(X_n=i)$$

which implies

$$\mathbb{P}^*(X_m = k) = \sum_{i=k}^{n-m+k} \binom{n-m}{i-k}\mathbb{P}^*(X_n = i) \tag{7}$$

and obviously (7) $\Rightarrow$ (8)

$$\mathbb{P}(X_m = k) = \sum_{i=k}^{n-m+k} \binom{n-m}{i-k}\binom{m}{k}\mathbb{P}^*(X_n = i) \tag{8}$$

By taking $m = 1$ in both instances in equations (3) and (8) we obtain the same probability of node $s_i$ being directly infected by node $v$ for all $i$, $1 \leqslant i \leqslant m$.

*4.2. Time and space complexity analysis of the FastSIR algorithm*

Here, we examine the average case running time and space complexity of the FastSIR algorithm. For order of growth of average case running time algorithm analysis we use standard big-$O$ notation (asymptotic upper bound within a constant factor) [30].

**Proposition 4.1.** *The average case running time of the FastSIR algorithm $\overline{T_f}$ is equal to:*

$$\overline{T_f} = O(\mathbb{E}[X]\overline{k}),$$

*where $\mathbb{E}[X]$ denotes total expected number of infected nodes and $\overline{k}$ average degree in network.*

Proof. Let us start with one infected node and its $k$ (degree) susceptible neighbors. Since distribution of number of infected neighbors is precalculated and it is possible to access to data with $O(1)$ we can neglect that to overall cost. The first step is uniformly choosing a value for a cumulative distribution. Since the parameters of $p$, $q$ and $k$ are known we should find appropriate number of infected nodes $k_1$ for that realisation. From the fact that there are $k + 1$ possible values we can find $k_1$ in $log(k)$ steps using binary search algorithm. In the next step, a random sample of $k_1$ nodes should be chosen, that would be infected, from $k$ of them. For that operation, calculation cost is proportional to $min(k_1, k - k_1)$ [22]. In the last step, infection should be transmitted to $k_1$ neighboring nodes. So the calculation cost is proportional to $k_1$. The overall running time for one infected node $T_f^1$ and $k$ susceptible neighbors can be calculated from the sum of costs for all three steps and it is equal to

$$T_f^1 = c_1 log(k) + c_2 min(k_1, k - k_1) + c_3 k_1 = O(k) \qquad (9)$$

where $c_1$, $c_2$ and $c_3$ are constants. Since $k_1 < k$, using big O notation it can be seen that average case running time for one node is $O(k)$. Hence, it does not depend on $1/q$. Total average case running time $\overline{T_f}$ is the sum of average times $\overline{T_f^i}$ for all infected nodes $v_i$ because main while loop of the FastSIR algorithm (see Algorithm 2) executes sequentially. This sum $\overline{T_f^1} + \overline{T_f^2} + ... + \overline{T_f^n}$ has $\mathbb{E}[X]$ terms which have $O(k)$ average case running time. Therefore average case running time $\overline{T_f}$ is equal to the O ( $\mathbb{E}[X]\overline{k}$ ) .

If asymptotic running times of FastSIR and Naive algorithm are compared, it can be seen that asymptotic upper bound average case running time of FastSIR is $1/q$ times lower. However, if we look in more detail in the running time for the case when $q$ equals one and we have an infected node and $k$ of its neighbors, their asymptotic upper bounds of average running times are the same. But, it can be seen that there is a run-time difference. Let that running times ratio between FastSIR and Naive SIR algorithms be $rq$ where $r$ is a value that depends on epidemic parameters and the network structure. Let that $k$ be big enough. In the case when $q$ equals one we have $r^1$. For FastSIR running time is equal to the sum in equation (9). For the Naive SIR algorithm that running time is $T_c^1 = c_4 k$ and $r^1 = T_f^1/T_c^1$. For FastSIR the first part of running time is from finding appropriate $k_1$ value for obtained random value of the cumulative distribution, the second part is from random selection of $k_1$ neighbors, while the last part is from the process of transmission of infection to $k_1$ neighbors. Since, the code for the last part of the FastSIR algorithm is almost equal to code of the Naive SIR algorithm, we can take that $c_4$ and $c_3$ are approximately equal. Whereas $k_1 \ll k$ for very small values of parameter $p$, it is expected that $r^1 < 1$. For some middle range of $p$ values $k_1$ is around $k/2$. For that case, it can be seen that $r^1$ can be greater than 1 if $c_2 > c_3$. When $p$ is near 1, $k_1 \approx k$, so the middle element of the sum is neglected and $r^1$ value is around 1 or slightly greater.

The value of $k_1$ is also dependent on $q$ parameter. In the case of same $p$, for smaller value of $q$, $k_1$ would be larger and vice versa. We can conclude that worst influence of epidemic parameters for duration of execution of the FastSIR algorithm would have $p$ and $q$ values for which $k_1 \approx k/2$.

The network structure has an influence on the ratio too. Looking at the for loop part of both algorithms we can see branching that depends of the state (susceptible or not susceptible) of the neighbouring node. Although upper bound values of $c_3$ and $c_4$ constants can be easily determined, their true values depend on the network structure. If structure has form of a tree or a chain, the infection can be transmitted only from one direction. The values of $c_3$ and $c_4$ are half the value and over half the value of upper bound in the cases of chain and tree, respectively. When the network structure has a lot of cycles, the infection can be transmitted from many directions and values of $c_3$ and $c_4$ are lower. Also, it is important to emphasise that $c_2$ is independent on the network structure. In accordance with above analysis the FastSIR algorithm would be slower than Naive Algorithm, for some specific values of epidemic parameters, if and only if $c_2 > c_3$. Hence, for the networks with more cycles there is a greater chance that, for some values of infected parameters, the Naive SIR algorithm would be faster.

The space complexity $S$ of the FastSIR algorithm with the respect to the number of links $L$, the number of nodes $N$ and the sum of all distinct degrees in network $K$ is equal to:

$$S[L, N, K] \approx \underbrace{2L}_{G} + \underbrace{K}_{C} + \underbrace{N}_{I} + \underbrace{N}_{S(v)} + \underbrace{N}_{R(v)} = 2L + K + 3N,$$

where the first term denotes space complexity of contact network $G$ (adjacency list), the second term denotes space complexity of cumulative distributions $C$ for all distinct degrees $k_i$ in $G$, the third term denotes the space complexity of a queue of infected nodes $I$, the next term denotes the space complexity of a vector indicator of susceptible nodes $S(v)$ and the last term denotes the space complexity of a vector indicator of recovered nodes $R(v)$. Note, that the $S(v)$ and $R(v)$ can be implemented as a bitset structure to further reduce memory consumption.

In connected networks $L \geqslant N$ and $2L \geqslant K$ and then the space complexity $S$ of the FastSIR algorithm is:
$$S[L, N, K] = O(L).$$
The values of $L$, $N$ and $K$ for studied networks are presented in Table 5.

### 4.3. Implementation of distribution precalculation

Looking at the cumulative distribution formula it can be seen that calculation cost is proportional to $k_{max}^4$. Speed up can be achieved using the fact that binomial coefficient values and the fraction part of the formula are repeated so caching them we can obtain lower calculation costs. However, we achieved further speed up using a recursive formula 10.

**Proposition 4.1.** *For each $k \neq 0$*

$$\mathbb{P}(X_n = k) = \frac{n}{k}\mathbb{P}(X_{n-1} = k-1) - \frac{n-k+1}{k}\mathbb{P}(X_n = k-1) \tag{10}$$

PROOF.

$$\mathbb{P}(X_n = k) = q\binom{n}{k}\sum_{l=0}^{k}\binom{k}{l}\frac{(-1)^l(1-p)^{n-k+l}}{1-(1-q)(1-p)^{n-k+l}}$$

$$= q\binom{n}{k}\sum_{l=0}^{k}\left[\binom{k-1}{l-1}+\binom{k-1}{l}\right]\frac{(-1)^l(1-p)^{n-k+l}}{1-(1-q)(1-p)^{n-k+l}}$$

$$= \underbrace{q\binom{n}{k}\sum_{l=0}^{k}\binom{k-1}{l-1}\frac{(-1)^l(1-p)^{n-k+l}}{1-(1-q)(1-p)^{n-k+l}}}_{=:S_1} + \underbrace{q\binom{n}{k}\sum_{l=0}^{k}\binom{k-1}{l}\frac{(-1)^l(1-p)^{n-k+l}}{1-(1-q)(1-p)^{n-k+l}}}_{=:S_2}$$

$$S_1 = -\frac{n-k+1}{k}q\binom{n}{k-1}\sum_{l=0}^{k-1}\binom{k-1}{l}\frac{(-1)^l(1-p)^{n-(k-1)+l}}{1-(1-q)(1-p)^{n-(k-1)+l}} =$$

$$= -\frac{n-k+1}{k}\mathbb{P}(X_n = k-1)$$

$$S_2 = q\binom{n}{k}\sum_{l=0}^{k-1}\binom{k-1}{l}\frac{(-1)^l(1-p)^{n-k+l}}{1-(1-q)(1-p)^{n-k+l}}$$

$$= \frac{n}{k}q\binom{n-1}{k-1}\sum_{l=0}^{k-1}\binom{k-1}{l}\frac{(-1)^l(1-p)^{(n-1)-(k-1)+l}}{1-(1-q)(1-p)^{(n-1)-(k-1)+l}}$$

$$= \frac{n}{k}\mathbb{P}(X_{n-1} = k-1) \Rightarrow \mathbb{P}(X_n = k) = \frac{n}{k}\mathbb{P}(X_{n-1} = k-1) - \frac{n-k+1}{k}\mathbb{P}(X_n = k-1)$$

Using this recursive formula the computation cost is proportional to $k_{max}^2$. It is very important to mention that in programming of cumulative distribution one should be very careful with precision. Because of that we use multiple precision library for this calculation. Empirically we obtained that it is safe to set the precision to be at least 0.8 times degree bits. The minimum precision is 64 bits. During the testing of calculation time we noticed that cost for large degree values predominantly depended on the precision used. The cumulative distribution values should be precalculated for a specific maximum degree only once and they can be used for all networks that have degrees less then maximum one. We consider that 50000 is high enough value of degree for majority of networks. Similar recursive formula can be used when random variable of time of recovery for each node is distributed as negative binomial probability distribution.

*4.4. Parallelization of the algorithm*

Like in similar algorithms [16], parallelization can be performed by partition of networks using MPI. Since we used a large number of repetitions it can be also naively parallelized performing each repetition on a separate core. The precalculation of distribution is also naturally parallelizable. Parallelization using GPUs is a very challenging task and it will be the scope of one of our next investigations.

## 5. Experimental results

In this section, we describe some detailed performance profiling and analysis of FastSIR implementation on our test server. The server has 4 Quad Core 2.4 GHz Intel E5330 processors and 50 GB of RAM memory. For test purposes we use only one core for each test. Algorithms are implemented in C using igraph [23] and gmp libraries [24].

The analysis was performed on several empirical networks: a network of 2003 condensed matter collaborations (cond-mat 2003) introduced in [4], an undirected, unweighted network representing the topology of the US Western States Power Grid (power grid) [8], a network of coauthorships between scientists posting preprints on the Astrophysics E-Print Archive between Jan 1, 1995 and December 31, 1999 (astro physics) [4], a symmetrized snapshot of the structure of the Internet at the level of autonomous systems, reconstructed from BGP tables posted by the University of Oregon Route Views Project (Internet) [27] and a network of Live Journal users (Live Journal) [29]. Table 1 shows the basic information for above mentioned networks.

Table 1: Basic network parameters

| Network | no of nodes | no of links | $k_{max}$ | $\bar{k}$ | sum of distinct degrees |
|---|---|---|---|---|---|
| Power grid | 4 941 | 6 594 | 19 | 2.7 | 142 |
| Cond-mat 2003 | 27 519 | 116 181 | 202 | 8.4 | 8 619 |
| Astro physics | 14 845 | 119 652 | 360 | 16.1 | 16 737 |
| Internet | 22 963 | 48 436 | 2390 | 4.2 | 32 118 |
| Live Journal | 5 189 809 | 77 365 447 | 15 023 | 29.6 | 2 503 563 |

For each analysis we measured running time of the Naive SIR algorithm and the FastSIR algorithm. Loading network structure data (adjacency list) from disc was not measured in running time analysis for both algorithms. However, loading precalculated probability distributions from disc was measured in running time analysis for the FastSIR algorithm. Also, we measured execution time for distribution precalculation. We studied the entire $(p, q)$ parametric space of the SIR model: a $[0, 1] \times [0, 1]$ square. The step value for both p and q was 0.1. Each simulation was started from the same node, and it was performed 2000 times. Upper bound memory consumption for all experiments was 9 GB. Although some authors use only a several dozen of repetitions, we consider that is not enough for stable results in the bimodal part of the phase space. The results of running time for $p$ values of 0.2, 0.5 and 0.8 and $q$ value of 0.1 for all tested networks are presented in Table 2. In addition in Table 3 are presented results for $p$ values of 0.2, 0.5 and 0.8 and $q$ values between 0.1 and 1. Graphs of results obtained for $p$ values of 0.2, 0.5 and 0.8 and different values of $q$ for all networks are presented in Figure 1, Figure 2 and Figure 3, respectively. Those figures show ratio of running time between Naive SIR and FastSIR.

Table 2: Running time in seconds for 2000 simulations, p = 0.2, 0.5, 0.8 and q = 0.1

| Network | p=0.2 | | p=0.5 | | p=0.8 | |
|---|---|---|---|---|---|---|
| | Naive SIR | FastSIR | Naive SIR | FastSIR | Naive SIR | FastSIR |
| Power grid | 3.2 | 0.4 | 7.0 | 0.9 | 7.1 | 0.9 |
| Cond-mat 2003 | 67.7 | 9.6 | 63.3 | 8.6 | 61.2 | 7.9 |
| Astro Physics | 44.1 | 7.0 | 41.2 | 5.9 | 39.9 | 5.1 |
| Internet | 42.5 | 5.0 | 41.6 | 4.9 | 40.5 | 4.7 |
| Live Journal | 50 683 | 6 699 | 48 373 | 5 635 | 47 531 | 5 078 |

Table 3: Running time in seconds for Live Journal network. Parameters: 2000 simulations, p = 0.2, 0.5 and 0.8, q = 0.1 to 1

| q | p=0.2 | | p=0.5 | | p=0.8 | |
|---|---|---|---|---|---|---|
| | Naive SIR | fastSIR | Naive SIR | fastSIR | Naive SIR | fastSIR |
| 0.1 | 50 683 | 6 699 | 48 373 | 5 635 | 47 531 | 5 078 |
| 0.2 | 25 841 | 7 200 | 24 398 | 6 314 | 24 067 | 5 357 |
| 0.3 | 18 550 | 7 200 | 16 580 | 6 841 | 16 276 | 5 609 |
| 0.4 | 13 686 | 6 987 | 12 870 | 7 259 | 12 329 | 5 843 |
| 0.5 | 13 197 | 6 704 | 10 394 | 7 591 | 9 951 | 6 060 |
| 0.6 | 9 394 | 6 400 | 8 720 | 7 859 | 8 345 | 6 253 |
| 0.7 | 8 301 | 6 073 | 7 513 | 8 072 | 7 185 | 6 429 |
| 0.8 | 8 744 | 5 805 | 6 622 | 8 250 | 6 293 | 6 592 |
| 0.9 | 7 521 | 5 508 | 5 869 | 8 555 | 5 597 | 6 749 |
| 1 | 5 291 | 5 259 | 5 064 | 8 666 | 5 082 | 7 777 |

Results differ between networks, but the trend is that the ratio is approximately proportional to $1/q$. When $q$ value is near one, the running time ratio differs dependingly on the network and the value of $p$. In can been seen that results are in accordance with the above analysis. When $p$ is small ($p = 0.2$), the FastSIR algorithm is faster or equal to Naive SIR for all $q$ values. But, when $p$ has value of 0.5, the Naive SIR algorithm is faster for larger q values for almost all networks. In addition when $p$ has value 0.8 the Naive SIR algorithm can be faster only for some networks and $q$ values very close to one. However, for small $q$ values the FastSIR algorithm is much faster and i.e. for $q$ value of 0.1, the ratio is between 7 and 9.5 depending on the network and the value of $p$.

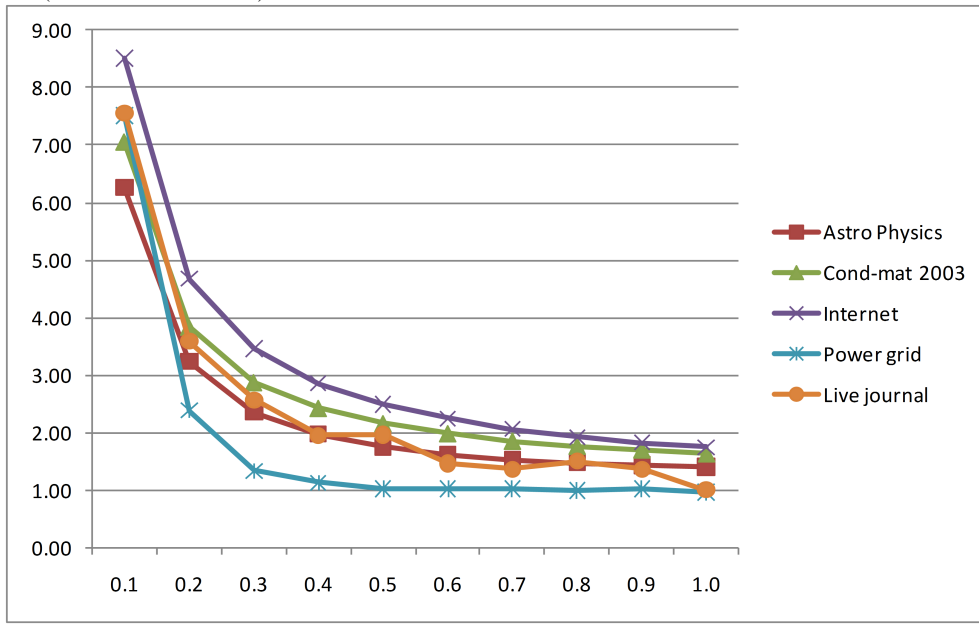

Figure 1: Running time ratio for Naive SIR and the FastSIR algorithm. Parameters: 2000 simulations, p = 0.2, q = 0.1 to 1 (Online version in colour).

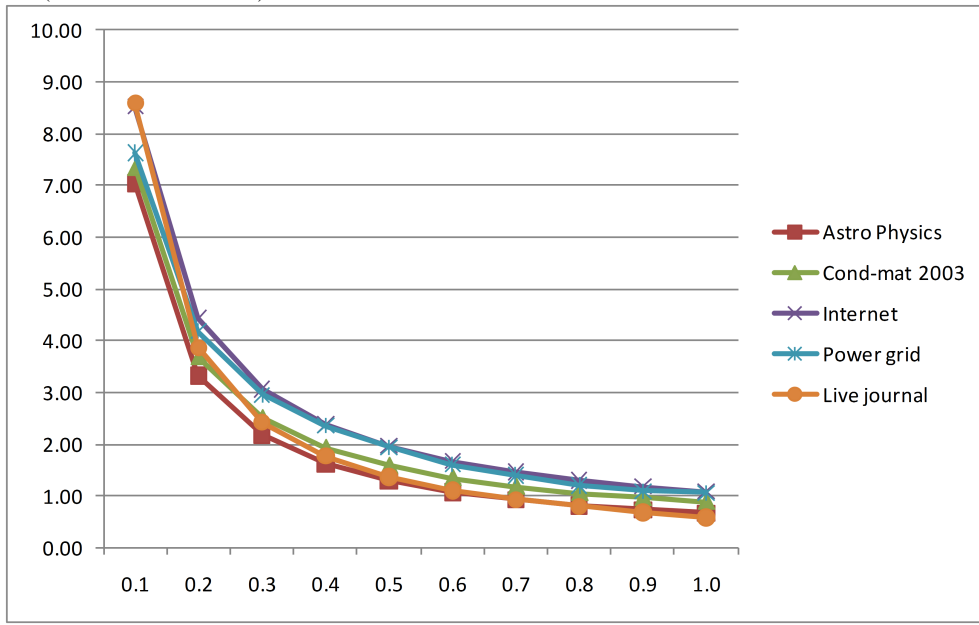

Figure 2: Running time ratio for Naive SIR and the FastSIR algorithm. Parameters: 2000 simulations, p = 0.5, q = 0.1 to 1 (Online version in colour).

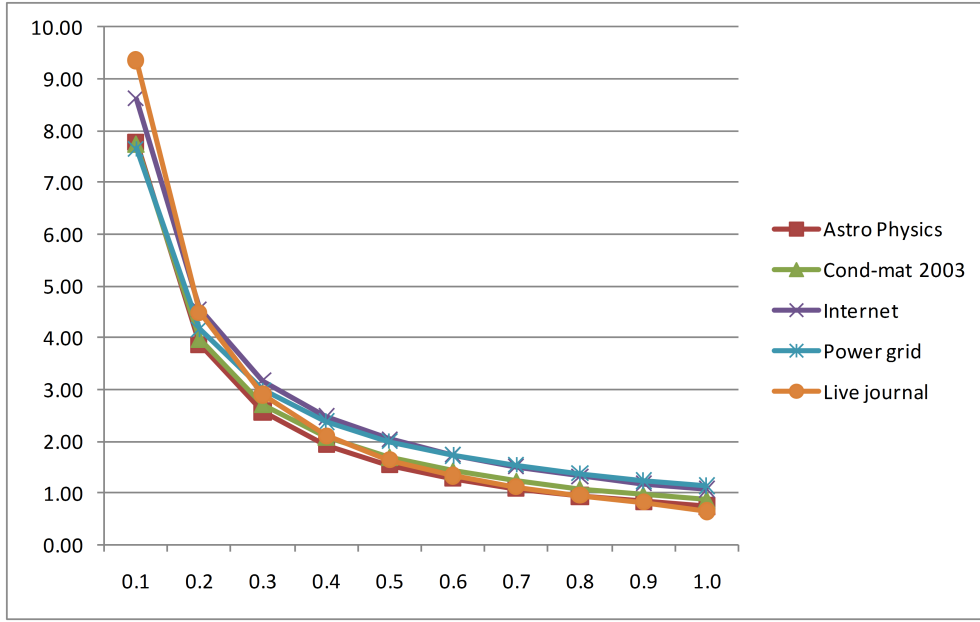

Figure 3: Running time ratio for Naive SIR and the FastSIR algorithm. Parameters: 2000 simulations, p = 0.8, q = 0.1 to 1 (Online version in colour).

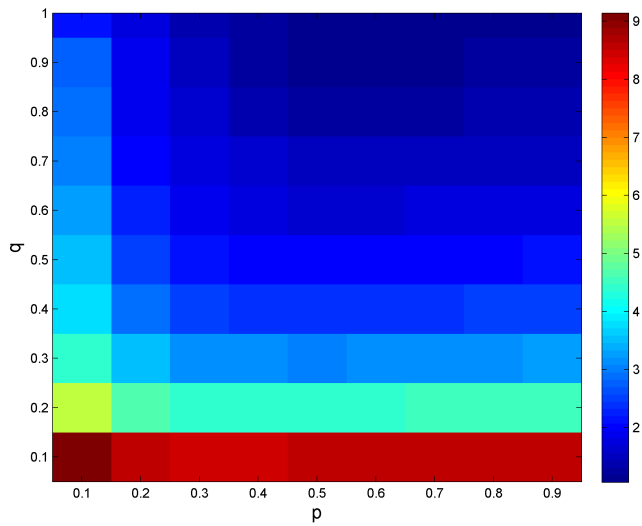

Figure 4: Running time ratio for Naive SIR and the FastSIR algorithm on parametric space on Internet network (Online version in colour).

Figure 5: Running time ratio for Naive SIR and the FastSIR algorithm on parametric space on Astro Physics network, white line represents border of the area of the parametric space where running time ratio is strictly less than one (Online version in colour).

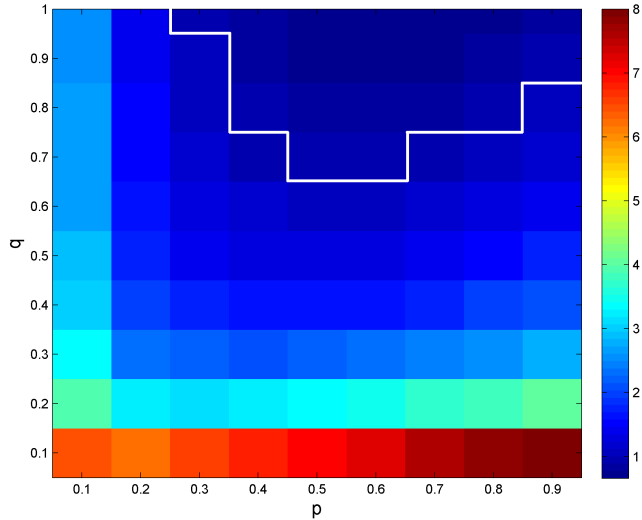

Above analysis of the running time ratio between Naive SIR and the FastSIR algorithm can be summarised in Figures 4, 5, 6, 7 and 8. In these Figures we show running time ratio on entire parametric space $(p, q)$ (averaged over 2000 simulations) and denote area of the parametric space (white line) where running time ratio is strictly less than one. It can also be seen that speed-up is dependent on the network structure. In accordance with analysis in Section 4, Naive algorithm is rarely faster than the FastSIR algorithm for networks of which structure is tree-like (Internet) or chain (Power grid). For networks with more cycles (i.e. Live Journal) Naive Algorithm is faster for more epidemic parameters.

It is very important to emphasise that results for Live Journal network of 5 mil. nodes and 77 mil. links are very fast. Average case running time for one simulation for p of 0.2 was between 2 and 4 seconds for the FastSIR algorithm. The worst obtained result for entire $(p, q)$ space was 5 seconds. Even results for the Naive SIR algorithm for that network did not exceed 30 seconds for one simulation. Furthermore, it should be stressed that results are achieved without parallelization. Hence, implementations of both algorithms can be used for large networks.

## 6. Application

Our version of the Naive SIR algorithm models full epidemic spreading dynamics and has low average case running time. Therefore, the Naive SIR algorithm is a core algorithm that models epidemic spreading and can easily be upgraded to parallel version with interventions. Interventions represent all sort of measures that are done by purpose in order to influence impact of epidemic spreading like using antiviral drugs or containment measures. This algorithm has wide areas of application in virus propagation, epidemic or knowledge spreading with SIR model. The FastSIR algorithm uses precalculated probability distributions of the number of infected nodes to reduce average case running time by a constant factor. The FastSIR algorithm has been used in study of the influence of the initially infected node to epidemic impacts due to its speed up when all nodes in network should be examined as initially infected node to calculate epidemic risks over entire parametric $(p, q)$ space [28]. This also has practical importance since the choice of the initially infected node may describe difference between a random outbreak and a terrorist attack.

Figure 6: Running time ratio for Naive SIR and the FastSIR algorithm on parametric space Cond-mat 2003 network, white line represents border of the area of the parametric space where running time ratio is strictly less than one (Online version in colour).

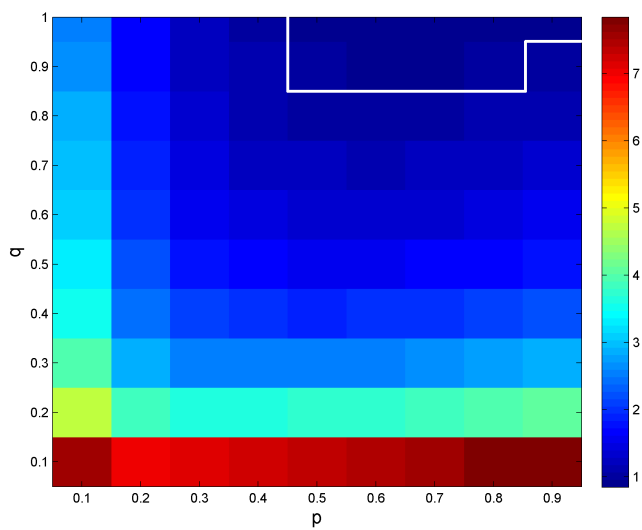

Figure 7: Running time ratio for Naive SIR and the FastSIR algorithm on parametric space on Live Journal network, white line represents border of the area of the parametric space where running time ratio is strictly less than one (Online version in colour).

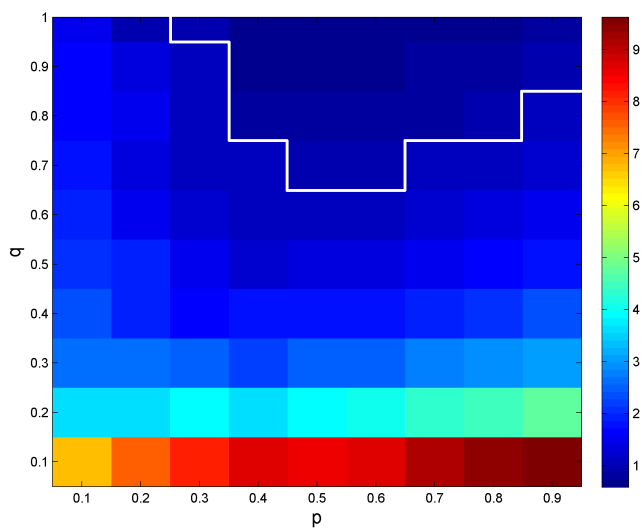

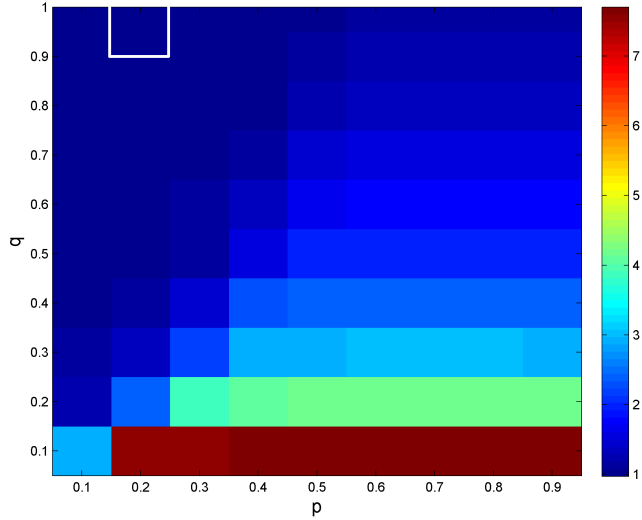

Figure 8: Running time ratio for Naive SIR and the FastSIR algorithm on parametric space on Power grid network, white line represents border of the area of the parametric space where running time ratio is strictly less than one (Online version in colour).

## 7. Discussion and conclusions

In this paper we have described how to construct our version of the Naive SIR algorithm and we have proposed completely new FastSIR algorithm for epidemic spreading simulations (SIR model) on an arbitrary network structure with reduced running time. Running time of the Naive SIR algorithm was reduced by using two data structures (queue and array) simultaneously for storing node states. This algorithm models full epidemic dynamics and thus can be easily upgraded to a parallel version with interventions (antiviral drugs and containments). Upper bound of the average case running time of the Naive SIR algorithm on m-arry tree has been analytically derived.

We propose a novel FastSIR algorithm to reduce the average case running time of the Naive SIR algorithm by approximately constant factor $1/q$ in one huge area of parametric space $(p, q)$. We also showed that average case running time of the FastSIR algorithm is equal to total expected number of infected nodes times average node degree. This low average case running time was accomplished by using precalculated probability distributions of the number of infected nodes along with binary search and a simple sampling algorithm. Correctness of this new algorithm has also been proven in four steps of equality of two algorithms. Precalculation of probability distributions of the number of infected nodes should be done with caution to avoid numerical errors. It is very important to mention that the FastSIR algorithm can be used in the same manner when time od recovery for each infected node is distributed as the negative binomial probability distribution.

It should also be clear that Naive SIR and FastSIR algorithms can be merged to a Hybrid SIR algorithm due to four steps of equivalence between Naive SIR and FastSIR algorithms. In one segment of the phase diagram Naive SIR can be faster than Fast SIR algorithm. Therefore Hybrid SIR algorithm could use one of two algorithms depending on position of an infective agent in the phase diagram in order to decrease running time. For the segment of the phase diagram where it is expected that Naive can be faster, we can estimate duration of both algorithm for a few simulations and then use the faster one for all other simulations (adaptive control).

Experimental analysis was made on five different empirical complex network on a single core processor. To the best of our knowledge our algorithms have a far shorter total running time per epidemic simulation than other algorithms. Parallelization of this algorithms was not in scope of this research and was left for future work. Empirical results show that the FastSIR algorithm is approximately faster than the Naive SIR algorithm by $1/q$ constant factor for a great part of parametric space $(p, q)$ which is in the excellent agreement with our expectations.

## Acknowledgements

The work of M.Š. is financed by Ministry of Education Science and Sports of the Republic of Croatia under Contract No. 036-0362214-1987, 098-1191344-2860 and by BMRC of A*STAR, Republic of Singapore. The work of H.Š. is supported by the Ministry of Education Science and Sports of the Republic of Croatia under Contract No. 098-0352828-2863.